\documentclass[11pt,english,epsf]{article}
\usepackage{lmodern}
\usepackage[T1]{fontenc}
\usepackage[latin9]{inputenc}
\usepackage{geometry}
\usepackage{amsmath}
\usepackage{amssymb}
\usepackage{graphicx}
\newtheorem{theorem}{Theorem}

\newcommand {\dfn} {\stackrel{\Delta} {=}}

\newcommand {\bu} {\mbox{\boldmath $u$}}

\newcommand {\bx} {\mbox{\boldmath $x$}}
\newcommand {\by} {\mbox{\boldmath $y$}}

\newcommand {\bE} {\mbox{\boldmath $E$}}

\newcommand{\calB}{{\cal B}}

\newcommand{\calE}{{\cal E}}

\newcommand{\calI}{{\cal I}}

\newcommand{\calQ}{{\cal Q}}

\newcommand{\calT}{{\cal T}}
\newcommand{\calU}{{\cal U}}

\newcommand{\calX}{{\cal X}}
\newcommand{\calY}{{\cal Y}}
\newcommand{\calZ}{{\cal Z}}

\begin{document}
\thispagestyle{empty}
\title{Guessing Individual Sequences: Generating Randomized Guesses Using
Finite--State Machines}
\author{Neri Merhav
}
\date{}
\maketitle

\begin{center}
The Andrew \& Erna Viterbi Faculty of Electrical Engineering\\
Technion - Israel Institute of Technology \\
Technion City, Haifa 32000, ISRAEL \\
E--mail: {\tt merhav@ee.technion.ac.il}\\
\end{center}
\vspace{1.5\baselineskip}
\setlength{\baselineskip}{1.5\baselineskip}

\begin{abstract}
Motivated by earlier results on universal randomized guessing, we consider an
individual--sequence approach to the guessing problem: in this setting, the
goal is to guess a secret, individual (deterministic) vector $x^n=(x_1,\ldots,x_n)$,
by using a finite--state machine that sequentially generates randomized
guesses from a stream of purely random bits. We define the finite--state
guessing exponent as the asymptotic normalized logarithm of the minimum
achievable moment of the number of 
randomized guesses, generated by any finite--state
machine, until $x^n$ is guessed successfully.
We show that the finite--state guessing exponent of any sequence is intimately
related to its
finite--state compressibility (due to Lempel and Ziv), and it is
asymptotically achieved by the decoder of 
(a certain modified version of) the 1978 Lempel--Ziv
data compression algorithm (a.k.a.\ the LZ78 algorithm), fed by purely random bits.
The results are also extended to the case where the guessing machine has
access to a side information sequence, $y^n=(y_1,\ldots,y_n)$, which is also
an individual sequence.\\

\end{abstract}

\newpage
\section{Introduction}

Consider the problem of guessing the realization of a finite--alphabet
random vector $X^n=(X_1,\ldots,X_n)$ using a series of yes/no questions of the form:
``Is $X^n=x^n(1)$?'', ``Is $X^n=x^n(2)$?'', and so on, until a positive
response is received. Given a distribution on $X^n$, a commonly used
performance metric for the guessing problem
is the expected number of guessing trials
required until $X^n$ is guessed correctly, or more generally, a general moment
of this number.

The design of \emph{guessing strategies} with the quest of minimizing
the moments of the number of guesses
has several applications in
information theory and related fields. One of them, for example, is
sequential decoding, as shown by Arikan \cite{Arikan96}, who built on the pioneering earlier work
of Massey \cite{Massey94} and related the best achievable guessing moment to the
R\'enyi entropy. More
modern applications of the guessing problem evolve around aspects of information security,
in particular, guessing passwords or decrypting messages protected by random keys.
For example, one may submit a sequence of guessing queries in attempt
to crack passwords -- see, e.g., \cite[Introduction]{MC19} (as well as
\cite{SHBCM19} and other references
therein) for a brief, yet quite comprehensive review
on guessing and security, as well as for some general 
historical overview of prior work
on the guessing problem in its large plethora of variants and extensions.

One of the main results in \cite{SHBCM19} is about devising optimal \emph{randomized
guessing} strategies, where instead of designing 
a particular, deterministic guessing list in advance,
one randomly draws independent guesses according to a carefully chosen
probability distribution of $n$--vectors, 
with the motivation that such a randomized strategy saves the
need of storing in memory long guessing lists and it also saves the need for
synchronization between the guesses of two or more agents who attempt to crack
the same password from different IP addresses in parallel. It turns out 
that with a clever choice of the randomized guessing distribution,
the resulting moment of the number of guesses (w.r.t.\ the randomness of both
$X^n$ and the guesses themselves) is exponentially the same as that of the
optimal deterministic guessing strategy. In a later work \cite{MC19},
this finding was further strengthened in two different ways: first, the framework
was extended from that of a discrete memoryless source (that governs $X^n$) 
to the much more general non--unifilar,
finite--state source (hidden Markov source model). Secondly, a universal
randomized guessing distribution was proposed, which is independent of the
unknown parameters of this finite--state source, as well as
the moment order of the number of guesses. The universal random guessing
distribution, $P(x^n)$, proposed in \cite{MC19}, was proportional to
$2^{-LZ(x^n)}$, where $LZ(x^n)$ designates the length (in bits) of the compressed
version of $x^n$ according to the LZ78 data compression algorithm \cite{ZL78}. Moreover,
practical algorithms for efficiently implementing this distribution were proposed in
\cite{MC19}. Finally, these results were also extended to account for the availability of
side information, $Y^n$, that is correlated to $X^n$, under a probabilistic
model where the sequence of pairs
$\{(X_i,Y_i)\}$ emerges from a non--unifilar finite--state source.

Motivated by those results of \cite{MC19}, in this work, we make an additional
step towards generality, by completely dropping the probabilistic assumption 
concerning the $n$--vector
to be guessed. In particular, we assume that it is an individual (deterministic),
finite--alphabet vector, which will be denoted by $x^n$. We also assume
that the independent random guesses are generated sequentially, using a finite--state machine,
fed by a sequence of purely random bits.\footnote{The supply of random bits
is assumed unlimited.} Inspired by the individual--sequence approach to data
compression, pioneered by Ziv and Lempel,
\cite{ZL78}, and followed later in the context of other tasks, like gambling
\cite{Feder91}, prediction \cite{FMG92}, and encryption \cite{merhav13},
we define the \emph{finite--state guessing exponent} as 
the minimum asymptotic exponential rate of the expectation of a given power of the
number of guesses, that is
achievable by any finite--state machine, and propose a universal
randomized guessing scheme that asymptotically achieves the
finite--state guessing
exponent. Similarly as in \cite{Feder91} and \cite{merhav13} (but in contrast
to \cite{FMG92}), we show that the finite--state guessing exponent is very
intimately related to the finite--state compressibility of the sequence to be
guessed. 

The proposed achievability scheme is basically the same as in
\cite{MC19}, which was based on the simple idea of
feeding the LZ78 decoder by purely random bits. While such a scheme cannot be
realized using a
finite--state machine, a simple twist can be offered, similary as done in
\cite{ZL78} in the context of compression: by employing this randomized
guessing scheme repeatedly and
resetting its memory at the beginning of every new block 
(however long), it becomes implementable as a
finite--state machine, and it still achieves the finite--state guessing
exponent in the limit of an increasing number of states. We therefore prove
that the same achievability scheme as in \cite{MC19} is asymptotically
optimal, not only in the probabilistic setting, but also in the
individual--sequence setting that we define here. Finally, we outline how these findings 
extend (with a few twists) to the case where the guessing machine has access
to a (deterministic) side information sequence.

The outline of the paper is as follows.
In Section \ref{ps}, we formalize the problem setting and spell out the
objectives. In Section \ref{conv}, we assert the converse theorem and prove
it. In Section \ref{direct}, we present the achievability scheme and prove the
direct theorem. Finally, in Section \ref{si}, we outline the main
modifications needed in order to extend the model and the results to the case
where side information is available to the guessing machine.

\section{Problem Setting}
\label{ps}

A finite--state guessing machine (FSGM) is defined by a sixtuplet
$Q=(\calU,\calX,\calZ,\Delta,f,g)$,
where $\calU=\{0,1\}$ is the binary input alphabet,
$\calX$ is a finite output alphabet of size $\alpha$,
$\calZ$ is a finite set of states,
$\Delta:\calZ\to\{0,1,2,\ldots\}$ defines the number of input bits processed
at each state,
$f:\calZ\times\calU^*\to\calX$ is the
output function, and $g:\calZ\times\calU^*\to\calZ$ is the
next--state function, where $\calU^*$ is a set of variable--length binary strings.
When a binary sequence, $\bu=u_1,u_2,\ldots$, $u_i\in\calU$,
$i=1,2,\ldots$, drawn from the binary symmetric source (BSS), is fed
into an FSGM $Q$, it produces an output sequence
$\hat{\bx}^n=(\hat{x}_1,\hat{x}_2,\ldots,\hat{x}_n)\in\calX^n$, 
while passing through a
sequence of states,
$z_1,z_2,\ldots,z_n$, according to
the following recursive equations, implemented for $i=1,2,\ldots,n$,
\begin{eqnarray}
t_i&=&t_{i-1}+\Delta(z_i),~~~~~~t_0\dfn 0 \label{ti}\\
v_i&=&(u_{t_{i-1}+1},u_{t_{i-1}+2},\ldots,u_{t_i}), \label{vi}\\
\hat{x}_i&=&f(z_i,v_i), \label{xi}\\
z_{i+1}&=&g(z_i,v_i), \label{nextstate}
\end{eqnarray}
where, without loss of generality, $z_1$ is assumed to be some fixed member of $\calZ$.
An FSGM $Q$ with $s$ states, or an $s$--state guessing machine,
is one with $|\calZ|=s$.

In the guessing game between Alice and Bob, Alice has access to a certain secret
$n$--vector $x^n=(x_1,\ldots,x_n)\in\calX^n$, while Bob is unaware of $x^n$, but is equipped
with an FSGM $Q$. In each guessing round,
Bob activates $Q$ by feeding it
with a sequence of purely random bits, $u_1,u_2,\ldots$, until an output sequence of
length $n$, $\hat{x}^n=(\hat{x}_1,\hat{x}_2,\ldots,\hat{x}_n)$, is obtained,
which is then submitted to Alice as a guess. Alice in her turn compares
$\hat{x}^n$ to $x^n$ and
returns an affirmative response if they match, and a negative one if they do not.
In the former case, $x^n$ has been guessed successfully and the guessing
process terminates. 
In the latter case, a new guessing round takes place, using new, independent
random bits, and so on, until $x^n$ is guessed successfully. 

Let $G_Q(x^n)$ be the random number of guessing rounds
needed for $Q$ until success.
For a given $\zeta > 0$, define
\begin{equation}
\label{gammas}
\gamma_s(x^n)=\min_{Q\in\calQ(s)}\frac{\ln\bE\{[G_Q(x^n)]^\zeta\}}{n},
\end{equation}
where $\calQ(s)$ is the set of all FSGMs with no more than $s$ states.
In order to define asymptotics, consider an
infinite sequence $\bx=(x_1,x_2,\ldots)$, whose components take on values in
$\calX$. We define the
asymptotic $s$--state guessing exponent as
\begin{equation}
\gamma_s(\bx)=\limsup_{n\to\infty}\gamma_s(x^n),
\end{equation}
and finally, the finite--state guessing exponent of $\bx$ is defined as
\begin{equation}
\gamma(\bx)=\lim_{s\to\infty}\gamma_s(\bx).
\end{equation}
While the minimizing FSGM $Q^*$ of eq.\ (\ref{gammas}) depends, in general, on
$x^n$, our objective is to devise a 
universal, sequential, randomized guessing scheme that is independent of $x^n$, and
yet it asymptotically achieves
$\gamma_s(x^n)$ in the limit of large $n$, 
followed by the limit of large $s$, and therefore it achieves
also $\gamma(\bx)$. Moreover, this universal guessing
scheme will not depend on the moment order $\zeta$ either.

Two observations regarding the above defined model of the FSGM are in order.\\

\noindent
1.\  As can be seen in eqs.\ (\ref{ti})--(\ref{nextstate}), at each cycle $i$,
the FSGM processes $\Delta(z_i)$ new input bits, in other words, a number of
bits that depends solely on the current
state, $z_i$. We could have defined the model to be seemingly more
general, where the number of input bits depends, not only on $z_i$, but also
on a certain number of the next incoming input bits,
$u_{t_{i-1}+1},u_{t_{i-1}+2},\ldots$, in the following manner. Consider a
situation where for each state $z\in\calZ$, one defines a binary tree, $T(z)$,
and define $\Delta(z_i,u_{t_{i-1}+1},u_{t_{i-1}+2},\ldots)$ to be the number
of branches of the path from
the root of $T(z)$ down to the leaf pertaining to the trail associated
with $u_{t_{i-1}+1},u_{t_{i-1}+2},\ldots$. On the other hand, if one does not
care about ``wasting'' input bits, this model can be formalized in the above
defined framework by re--defining $\Delta(z)$ to be the length of the path from
the root to the deepest leaf and then extending $T(z)$ to be a full binary tree of
depth $\Delta(z)$, where $f$ and $g$ are defined to be the same for all
descendants of every given leaf of the original tree, $T(z)$. For example,
if $T(z)$ originally contains the three leaves, corresponding to the binary paths `0', `10' and `11', 
then we extend the path `0' to its two children,
`00' and `01' to obtain a full binary tree of four leaves and depth $\Delta(z)=2$, but we
let $f(z,00)=f(z,01)$ and
$g(z,00)=g(z,01)$ (see also the example in comment no.\ 2 below), 
so that the second bit after `0' is not really used and
hence is immaterial. 
Therefore, the only difference between the 
FSGM with the extended tree and the original FSGM is
that the second bit of `00' and `01' is ``wasted'',
but since we are assuming, in this paper, that resources of randomness are unlimited, this is
inconsequential. The reason we prefer the model of $\Delta(z_i)$ over the model of
$\Delta(z_i,u_{t_{i-1}+1},u_{t_{i-1}+2},\ldots)$ is just its relative
simplicity.\\

\begin{figure}[ht]
\vspace*{1cm}
\hspace*{5cm}\input{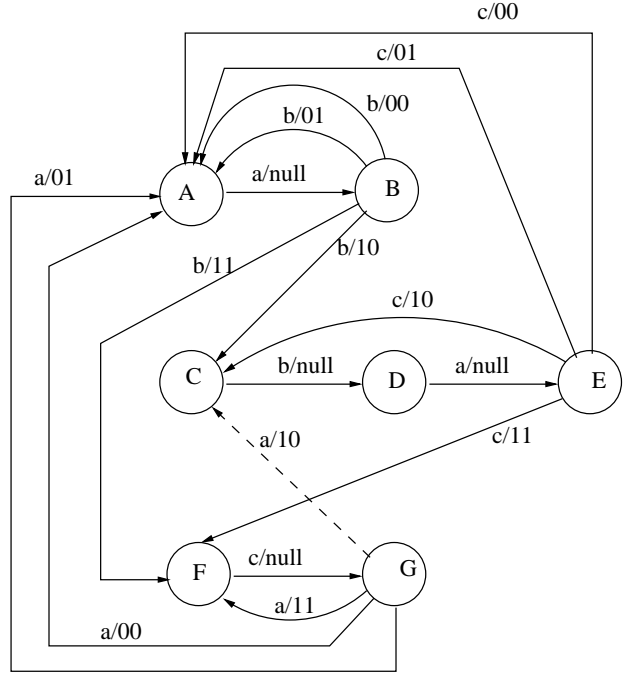}
\caption{\small State transition diagram of the variable--to--variable
length mapping (\ref{mapping}).}
\label{fsm}
\end{figure}

\noindent
2.\ The above defined model of the FSGM is general enough to operate as any
(state--dependent) mapping from variable--length binary
input strings to variable--length strings of symbols, such as decoders
corresponding to
variable--to--variable length encoders for data compression.
This is the case when we
allow $\Delta(z)=0$
at some states, as it enables the finite--state machine 
to idle between successive readings of chunks of
input bits. To implement such a variable--to--variable length mapping in the
framework of our model, the system works 
as follows. Upon receiving a variable--length binary input string $v_i$, the
system passes along a certain sequence of states, where at each state it
outputs one output symbol without reading any new input bits
($\Delta(z_j)=0$), until the entire output word is produced. As a simple
example, consider the variable--to--variable length mapping
\begin{equation}
\label{mapping}
0\to\mbox{ab}~~~~
10\to\mbox{bac}~~~~
11\to\mbox{ca}.
\end{equation}
The state transition diagram of the associated FSM is depicted in Fig.\
\ref{fsm}, as a graph whose vertices represent the states and whose edges
designate the state transitions.

Each state transition is defined by a pair $(z,v)$, $z\in\calZ$,
$v\in\calU^{\Delta(z)}$, and its edge is labeled in Fig.\ \ref{fsm} by the
corresponding output $\hat{x}=f(z,v)$, followed by a slash, and followed in
turn by the contents of $v$. For
instance, states and A and B represent the assignment $0\to \mbox{ab}$ or equivalently,
the pair of assignments $00\to \mbox{ab}$ and $01\to \mbox{ab}$ (see comment
no.\ 1 above). 
The transition from state A to state B is labeled by ``a/null'',
which means $\Delta(\mbox{A})=0$ and $v$ is null (no new bits are read by the
system), and the 
output is $\mbox{a}=f(\mbox{A},\mbox{null})$.
This means that after visiting at state A, the machine must pass to
state B (as $\mbox{B}=g(\mbox{A},\mbox{null})$), without processing input bits. 
From state B, all outgoing edges
are associated with the output ``b'', and $\Delta(\mbox{B})=2$ new bits 
are read for the next round of the mapping (\ref{mapping}):
if the first bit is `0', then we are back to state A, otherwise,
we pass to state C or state F, to map `10' or `11', respectively, and we proceed
similarly as before.

\section{Converse Theorem}
\label{conv}

\subsection{Preliminaries}

Before presenting the converse theorem and its proof, we need to recall the
notions of finite--state compressibility and string parsing from \cite{ZL78} (which
inspired the above definition of the FSGM).
To make the presentation self--contained, we provide the definitions of these
terms here, but with a slightly different notation, not to confuse with the
above notation of the FSGM.

A $K$--state
encoder $E$ is defined by a quintuple
$(\calX,\calB,\Sigma,f,g)$, where
$\calX$ is the
alphabet of size $\alpha$ of the source sequence to be compressed,
$\calB$ is
a finite set of binary words
(possibly of different lengths, including the null word for idling),
$\Sigma$ is a finite set
of $K$ states,
$p:\Sigma\times\calX\to\calB$ is the encoder output function, and
$q:\Sigma\times\calX\to\Sigma$ is the next--state function.
When an input sequence $(x_1,x_2,...)$ is
fed sequentially into $E$, the
encoder outputs a sequence of binary words $(b_1,b_2,\ldots)$, $b_i\in\calB$,
while going through a sequence of states $(\sigma_1,\sigma_2,\ldots)$,
$\sigma_i\in\Sigma$,
according to
\begin{equation}
\label{fsmcompression}
b_i=p(\sigma_i,x_i),~~~\sigma_{i+1}=q(\sigma_i,x_i),~~~i=1,2,...
\end{equation}
where $\sigma_i$ is the state of $E$ at time instant $i$
and where the initial state, $\sigma_1$, is assumed a fixed member of $\Sigma$.
The decoder receives the compressed bit--stream,
$b_1,b_2,\ldots$,
and reconstructs $x_1,x_2,\ldots$.
A finite--state encoder $E$
is said to be {\it information
lossless} (IL) if for all $\sigma_1\in\Sigma$ and all
$x^n\in\calX^n$, $n\ge 1$,
the triple $(\sigma_1,\sigma_{n+1},b^n)$ (with $b^n\dfn(b_1,\ldots,b_n)$) 
uniquely determines $x^n$,
where $\sigma_{n+1}$ and $b^n$ are obtained by iterating
eq.\ (\ref{fsmcompression}) with $\sigma_1$
and $x^n$ as inputs. The length function
associated with $E$ is defined as $L_E(b^n)=\sum_{i=1}^nl(b_i)$,
where $l(b_i)$ is the length of the binary string $b_i\in\calB$.
The compression ratio of $x^n$ w.r.t.\ $E$ is defined as\footnote{Note that
here, unlike in \cite{ZL78}, we define the compression ratio without
normalization by $\log\alpha$, where $\alpha$ is the source alphabet size.}
\begin{equation}
\rho_E(x^n)=\frac{L_E(b^n)}{n}.
\end{equation}
Next, we define
\begin{equation}
\rho_K(x^n)=\min_{E\in\calE(K)}\rho_E(x^n),
\end{equation}
where $\calE(K)$ is the class of all IL encoders with no more than $K$ states.
Furthermore, for an infinite sequence $\bx=(x_1,x_2,\ldots)$, let
\begin{equation}
\rho_K(\bx)=\limsup_{n\to\infty}\rho_K(x^n),
\end{equation}
and finally,
\begin{equation}
\rho(\bx)=\lim_{K\to\infty}\rho_K(\bx).
\end{equation}
The following converse theorem was asserted and proved in
\cite{ZL78} in the context of data compression.

\begin{theorem}\cite[Theorem 1]{ZL78}
For every $x^n\in\calX^n$,
\begin{equation}
\rho_K(x^n)\ge \frac{[c(x^n)+K^2]}{n}\log\left[\frac{c(x^n)+K^2}{4K^2}\right]+\frac{2K^2}{n},
\end{equation}
where $c(x^n)$ is the largest number of distinct strings (or phrases) whose concatenation
forms $x^n$.
\end{theorem}

\subsection{Main Result}

Returning to our guessing problem, our converse theorem relates the
finite--state guessing exponent of $x^n$ to its finite--state compressibility.

\begin{theorem}
\label{converse}
Let $Q$ be an arbitrary FSGM with $s$ states and let $x^n$ be a sequence
generated from $Q$ with probability less than $1/2$. Then,
\begin{eqnarray}
\gamma_s(x^n)&\ge&\zeta\cdot\max_{\{\ell~\mbox{divides}~n\}}\left[\rho_{K(\ell)}(x^n)-
\frac{\log(2s^3e)}{\ell}\right]-\frac{\log(e^22^\zeta)}{n};\\
\gamma_s(x^n)&\ge&\zeta\bigg[c(x^n)\ln c(x^n)-n\delta_n(s)\bigg],
\end{eqnarray}
where $K(\ell)\dfn(\alpha^{\ell+1}-1)/(\alpha-1)$ and
$\delta_n(s)$ tends to zero uniformly as $n\to\infty$ for any fixed $s$.
\end{theorem}

\noindent
{\it Discussion.} 
The condition that $x^n$ is generated from $Q$ with probability less than
$1/2$ is purely technical and it is almost trivially always met. 
It means that at least one input bit
is utilized for the generation of $x^n$. 
Theorem \ref{converse} provides 
two different lower bounds on $\gamma_s(x^n)$.
The first one relates the $s$--state guessing exponent with the $K$--state 
compressibility of $x^n$, and it
implies also an asymptotic version of our converse theorem, which is
$\gamma(\bx)\ge\zeta\cdot\rho(\bx)$. This lower bound will be shown to be achieved
asymptotically by a certain sequence of FSGMs whose number of
states goes to infinity {\it after} $n\to\infty$, thus establishing also the
reversed inequality $\gamma(\bx)\le\zeta\cdot\rho(\bx)$, and hence an
equality, 
\begin{equation}
\gamma(\bx)=\zeta\cdot\rho(\bx).
\end{equation}
The second lower bound, which is in fact a corollary of the first one, and
hence is weaker, has the advantage that it
is computable. It can be achieved by a randomized guessing
machine whose number of states is unlimited. The details are deferred to the
next section, where the achievability results will be asserted and proved.\\

\noindent
{\it Proof of Theorem \ref{converse}.} 
Let $Q$ be an arbitrary FSGM with $s$ states. First, observe that 
since the input bits are i.i.d., the output of $Q$
is a non--unifilar\footnote{A non--unifilar finite--state source is a
finite--state source where the
underlying state sequence is hidden, namely, it cannot be recovered from the
source sequence alone. A unifilar source is obtained as a special case, where
eq.\ (\ref{nextstate}) is replaced by the equation $z_{i+1}=g(z_i,\hat{x}_i)$.
Note that since $\hat{x}_i=f(z_i,v_i)$, then eventually, here too $z_{i+1}$ is
a function of $(z_i,v_i)$, but not every function of $(z_i,v_i)$ can be
represented as a composition of $f$ and $g$ in the form,
$z_{i+1}=g(z_i,f(z_i,v_i))$. Thus, our model is more general than the model of
unifilar finite--state sources.}
finite--state source, a.k.a.\ a hidden Markov source. In particular,
\begin{equation}
\label{hmm}
P(\hat{x}^n|z_1)=\sum_{z_2,z_3,\ldots,z_{n+1}}\prod_{i=1}^n P(\hat{x}_i,z_{i+1}|z_i),
\end{equation}
where $P(\hat{x},z^\prime|z)=m(\hat{x},z^\prime|z)\cdot 2^{-\Delta(z)}$,
$m(\hat{x},z^\prime|z)$ being the number of 
binary strings $\{v\}$ of length $\Delta(z)$ such that
$f(z,v)=\hat{x}$ and $g(z,v)=z^\prime$, as each such binary string
has probability $2^{-\Delta(z)}$.
Since $z_1$ is assumed fixed, the
conditioning on $z_1$ will henceforth be dropped. Now, since the guesses are
statistically mutually independent,
\begin{equation}
\bE\{[G_Q(x^n)]^\zeta\}=\sum_{k=1}^\infty
k^\zeta[1-P(x^n)]^{k-1}P(x^n).
\end{equation}
Consider now the following chain of inequalities, holding for any $q\in(0,1]$:
\begin{eqnarray}
\sum_{k=1}^\infty k^\zeta(1-q)^{k-1}\cdot q&\ge&
q\cdot\sum_{k=\lfloor 1/q\rfloor}^{\infty} k^\zeta(1-q)^{k-1}\nonumber\\
&\ge& q\cdot\sum_{k=\lfloor 1/q\rfloor}^{\infty}
\left(\bigg\lfloor\frac{1}{q}\bigg\rfloor\right)^\zeta(1-q)^k\nonumber\\
&\ge&q\cdot\left(\frac{1}{q}-1\right)^\zeta\cdot\frac{(1-q)^{\lfloor 1/q\rfloor}}
{1-(1-q)}\nonumber\\
&\ge&\left(\frac{1}{q}-1\right)^\zeta\cdot (1-q)^{1/q}\nonumber\\
&=&\left(\frac{1}{q}-1\right)^\zeta\cdot \exp\left\{\frac{1}{q}\ln(1-q)\right\}\nonumber\\
&\ge&\left(\frac{1-q}{q}\right)^\zeta\cdot \exp\left\{-\frac{1}{1-q}\right\},
\end{eqnarray}
where in the last step, we have used the inequality $\ln(1+t)\ge
\frac{t}{1+t}$, which holds for all $t > -1$.
Thus, with the assignment $q=P(x^n)$, and since it is assumed that
$P(x^n)\le 1/2$, we have
\begin{eqnarray}
\label{guess-prob}
\bE\{[G_Q(x^n)]^\zeta\}&\ge&
\frac{2^{-\zeta}}{e^2}\cdot[P(x^n)]^{-\zeta}\nonumber\\
&=&\frac{2^{-\zeta}}{e^2}\cdot \exp_2\{-\zeta\log P(x^n)\}.
\end{eqnarray}
Assume that $\ell$ divides $n$ and consider the partition of $\hat{x}^n$ into
$m=n/\ell$ non--overlapping blocks of length $\ell$, $\hat{x}^n=
(\hat{x}_1^{\ell},\hat{x}_{\ell+1}^{2\ell},\ldots,\hat{x}_{n-\ell+1}^n)$, 
where here and throughout the sequel, the notation $\hat{x}_i^j$ ($i<j$)
denotes the string segment $(\hat{x}_i,\hat{x}_{i+1},\ldots,\hat{x}_j)$ (and a
similiar notation applies also to other vectors).
We also define the diluted sequence of states
$z^m\dfn(z_1,z_{\ell+1},z_{2\ell+1},\ldots,z_{n-\ell+1})$. Then, it follows from
eq.\ (\ref{hmm}) that
\begin{equation}
P(\hat{x}^n,z^m)=\prod_{i=0}^{m-1}P(\hat{x}_{i\ell+1}^{i\ell+\ell},z_{i\ell+\ell+1}|z_{il+1}).
\end{equation}
Let $\calT(\hat{x}^n|z^m)$ denote the set of vectors in $\calX^n$ that are
obtained by permuting different $\ell$--blocks that begin at the
same state, $z$, and end at the same state, $z^\prime$. Obviously, all members
of $\calT(\hat{x}^n|z^m)$ have the same joint probability with $z^m$.
A simple combinatorial
argument (see, e.g., \cite[eq.\ (A.11)]{merhav91}) yields that
\begin{equation}
\log|\calT(\hat{x}^n|z^m)|\ge m[\hat{H}_\ell(x^n)-\log(s^2e)],
\end{equation}
where $\hat{H}_\ell(x^n)$ is the entropy associated with the
empirical distribution of the non--overlapping $\ell$--blocks of 
$\hat{x}^n$, that is,
\begin{equation}
\hat{H}_\ell(\hat{x}^n)=-\sum_{a^\ell\in\calX^\ell}\hat{P}(a^\ell)\log
\hat{P}(a^\ell),
\end{equation}
where
\begin{equation}
\hat{P}(a^\ell)=\frac{1}{m}\sum_{i=0}^{m-1}\calI\{\hat{x}_{i\ell+1}^{i\ell+\ell}=a^\ell\},~~~
a^\ell\in\calX^\ell
\end{equation}
$\calI\{\hat{x}_{i\ell+1}^{i\ell+\ell}=a^\ell\}$ being the indicator function for
the event $\{\hat{x}_{i\ell+1}^{i\ell+\ell}=a^\ell\}$.
Now, since
\begin{eqnarray}
1\ge\sum_{\tilde{x}^n\in\calT(\hat{x}^n|z^m)}P(\tilde{x}^n,z^m)=|\calT(\hat{x}^n|z^m)|\cdot
P(\hat{x}^n,z^m),
\end{eqnarray}
it follows that
\begin{eqnarray}
P(\hat{x}^n)&=&\sum_{z^m}P(\hat{x}^n,z^m)\nonumber\\
&\le&\sum_{z^m}\frac{1}{|\calT(\hat{x}^n|z^m)|}\nonumber\\
&\le&s^{m}\cdot
2^{-m[\hat{H}_\ell(x^n)-\log(s^2e)]}\nonumber\\
&=&2^{-m[\hat{H}_\ell(x^n)-\log(s^3e)]},
\end{eqnarray}
or, equivalently,
\begin{equation}
\label{prob-ent}
-\log P(\hat{x}^n)\ge m[\hat{H}_\ell(x^n)-\log(s^3e)].
\end{equation}
It remains to lower bound the r.h.s.\ in terms of the finite--state compressibility.
Consider a Shannon code w.r.t.\ an arbitrary probability distribution $F$
of $\ell$--vectors, that is, a code that assigns $\lceil-\log
F(x^\ell)\rceil$ bits to the lossless compression of $x^\ell$. 
Similarly as argued in \cite[p.\ 2245, right column]{merhav00}, such a code can be implemented
by an IL finite--state encoder with
$\sum_{j=0}^{\ell}\alpha^j=(\alpha^{\ell+1}-1)/(\alpha-1)=K(\ell)$
states, and so, by the definition of the $K$--state compressibility,
\begin{eqnarray}
n\rho_{K(\ell)}(\hat{x}^n)&\le&
\sum_{i=0}^{m-1}\lceil-\log F(\hat{x}_{i\ell+1}^{i\ell+\ell})\rceil\nonumber\\
&\le&-\sum_{i=0}^{m-1}\log F(\hat{x}_{i\ell+1}^{i\ell+\ell})+m,
\end{eqnarray}
and since this is true for every probability distribution $F$ on $\calX^\ell$, 
the r.h.s.\ may be minimized w.r.t.\
$F$, to obtain
\begin{equation}
n\rho_{K(\ell)}(\hat{x}^n)\le m\hat{H}_\ell(\hat{x}^n)+m.
\end{equation}
Combining this with eq.\ (\ref{prob-ent}), we obtain
\begin{equation}
-\log P(\hat{x}^n)\ge n\rho_{K(\ell)}(\hat{x}^n)-m\log(2s^3e),
\end{equation}
which, together with eq.\ (\ref{guess-prob}), 
proves the first inequality of the converse theorem, since $\ell$ is an
arbitrary divisor of $n$. 

Finally, the second lower bound of the converse theorem is obtained from
\cite[Theorem 1]{ZL78}, as follows.
\begin{eqnarray}
n\rho_{K(\ell)}(\hat{x}^n)&\ge&[c(\hat{x}^n)+K^2(\ell)]\log
\frac{c(\hat{x}^n)+K^2(\ell)}{4K^2(\ell)}\nonumber\\
&\ge& c(\hat{x}^n)\log c(\hat{x}^n)-[c(\hat{x}^n)+K^2(\ell)]\log[4K^2(\ell)]\nonumber\\
&\ge& c(\hat{x}^n)\log
c(\hat{x}^n)-\frac{n\log[4K^2(\ell)]\log\alpha}{(1-\epsilon_n)\log
n}-K^2(\ell)\log[4K^2(\ell)],
\end{eqnarray}
where $\epsilon_n\to 0$, and the last step follows from the inequality 
$c(x^n)\le n\log\alpha/[(1-\epsilon_n)\log n]$ \cite[Theorem 2]{LZ76},
\cite[Lemma 13.5.3, p.\ 450]{CT06}.
It follows that the second lower bound of Theorem \ref{converse} holds with
\begin{equation}
\delta_n(s)=\min_{\{\ell~\mbox{divides}~n\}}\left[\frac{\log[4K^2(\ell)]\log\alpha}
{(1-\epsilon_n)\log
n}+\frac{K^2(\ell)\log[4K^2(\ell)]}{n}+\frac{\log(2s^3e)}{\ell}\right]+\frac{\log(e^22^\zeta)}{n}.
\end{equation}
This completes the proof of Theorem \ref{converse}.

\section{Direct Theorem}
\label{direct}

\subsection{Preliminaries}

Before presenting our direct theorem (achievability), we need to recall a few
more terms and facts from \cite{ZL78}. 

The incremental parsing procedure of the LZ78 algorithm is a procedure of
sequentially parsing a vector $x^n$ such that each new phrase is the shortest
string that has not been encountered before as a parsed phrase, with the
possible exception of the last phrase, which might be incomplete. For example,
the incremental parsing of the vector $x^{15}=\mbox{abbabaabbaaabaa}$ is
$\mbox{a,b,ba,baa,bb,aa,ab,aa}$. Let $c_{\mbox{\tiny LZ}}(x^n)$ denote the
number of phrases in $x^n$ resulting from the incremental parsing procedure.
Obviously, $c_{\mbox{\tiny LZ}}(x^n)\le c(x^n)+1$ \cite[Theorem 2]{LZ76}, 
\cite[eq.\ (6)]{ZL78}, as $c(x^n)$ was defined
above as the maximum number of distinct phrases. Let $LZ(x^n)$ denote the
length of the LZ78 binary compressed code for $x^n$. According to \cite[Theorem 2]{ZL78},
\begin{eqnarray}
\label{lz-clogc}
LZ(x^n)&\le&[c(x^n)+1]\log\{2\alpha[c(x^n)+1]\}\nonumber\\
&=&c(x^n)\log[c(x^n)+1]+c(x^n)\log(2\alpha)+\log\{2\alpha[c(x^n)+1]\}\nonumber\\
&=&c(x^n)\log c(x^n)+c(x^n)\log\left[1+\frac{1}{c(x^n)}\right]+
c(x^n)\log(2\alpha)+\log\{2\alpha[c(x^n)+1]\}\nonumber\\
&\le&c(x^n)\log c(x^n)+\log
e+\frac{n(\log\alpha)\log(2\alpha)}{(1-\epsilon_n)\log n}+\log[2\alpha(n+1)]\nonumber\\
&\dfn&c(x^n)\log c(x^n)+n\cdot\epsilon(n),
\end{eqnarray}
where $\epsilon(n)$ clearly tends to zero as $n\to\infty$, at the rate of
$1/\log n$.

\subsection{Main Result}

Returning to the guessing problem,
our direct theorem is as follows.

\begin{theorem}
\label{directtheorem}
\begin{itemize}
\item [(a)] Consider the random guessing distribution
\begin{equation}
P(\hat{x}^n)=\frac{2^{-LZ(\hat{x}^n)}}{\sum_{x^n\in\calX^n}2^{-LZ(x^n)}}.
\end{equation}
Then,
\begin{equation}
\log \bE\{[G(x^n)]^\zeta\}\le \zeta\cdot c(x^n)\log c(x^n)+n\cdot O\left(\frac{1}{\log
n}\right).
\end{equation}
\item[(b)]
Let $\ell$ divide $n$ and consider the product form random guessing
distribution,
\begin{equation}
P(\hat{x}^n)=\prod_{i=0}^{n/\ell-1}\left[
\frac{2^{-LZ(\hat{x}_{i\ell+1}^{i\ell+l})}}{\sum_{x^\ell\in\calX^\ell}2^{-LZ(x^\ell)}}\right].
\end{equation}
Then, for every positive integer $K$,
\begin{equation}
\log \bE\{[G(x^n)]^\zeta\}\le \zeta
n\cdot\left[\rho_K(\hat{x}^n)+\frac{\log(4K^2)}{(1-\epsilon_\ell)\log\ell}+
\frac{K^2\log(4K^2)}{\ell}+\epsilon(\ell)\right].
\end{equation}
\end{itemize}
\end{theorem}

\noindent
{\it Discussion.} Part (a) of Theorem \ref{directtheorem} is an achievability
result that is matching the second lower bound in Theorem
\ref{converse}. However, this is incompatible with the framework of
finite--stage machines since this random guessing distribution 
cannot be implemented with a finite--state
machine as $n$ grows without bound. The random guessing distribution of part(b), 
on the other hand, can be
implemented using an FSGM with no more than $\ell\cdot\alpha^\ell$ states, and
it is a matching achievability result to the first lower bound of Theorem \ref{converse}.
It should be pointed that even the random guessing distribution of part (a) can be
implemented efficiently using practical algorithms, as described in
\cite{MC19}. The upper bound of part (b) is meaningful when
$K^2\ll\ell$.

Theorems \ref{converse} and \ref{directtheorem} together tell us that
essentially the best achievable guessing moment $\bE\{[G(x^n)]^\eta\}$ is
of the exponential order of $2^{\zeta c(x^n)\log c(x^n)}$. In general, when the $\eta$-th
moment of a random variable behaves like $A^\eta$ for some positive constant
$A$ that is independent of $\eta$, and every $\eta > 0$,
it indicates that this random variable is (nearly) degenerate. In other words, 
it concentrates very rapidly around
its mean. Indeed, it can easily be shown very similarly\footnote{In eq.\ (17)
of \cite{MC19}, this it is shown for the empirical entropy (instead
of $c(x^n)\log c(x^n)$, but the proof for $c(x^n)\log c(x^n)$ is exactly the
same.} as in
\cite[eq.\ (17)]{MC19}, that the probability of the
event $\{G(x^n)\ge 2^{c(x^n)\log c(x^n)+n\epsilon}\}$ decays double
exponentially rapidly for the optimal guessing distribution, provided that
$\epsilon > 0$.\\

\noindent
{\it Proof of Theorem \ref{directtheorem}.}
Part (a) is almost a restatement of \cite[Theorem 3]{MC19}. The proof is
therefore almost identical to the proof of that result.
The only difference is
that here, since $x^n$ is a given deterministic vector, the final step in
\cite[proof of Theorem 3]{MC19}, of taking the expectation w.r.t.\ the
randomness of $x^n$, is now omitted, and the expectation of
$[G_Q(x^n)]^\zeta$ is taken only w.r.t.\ the
randomness of the guesses, which as is shown in \cite[Lemma 1]{MC19}, behaves
like $[P(x^n)]^{-\zeta}$, where $P(\cdot)$ is the
random guessing distribution.

As for part(b) of Theorem \ref{directtheorem},
since $LZ(\cdot)$ is a uniquely decipherable code, it satisfies Kraft's
inequality, and so,
\begin{eqnarray}
P(\hat{x}^n)&=&\prod_{i=0}^{n/\ell-1}\frac{2^{-LZ(\hat{x}_{\ell
i+1}^{\ell i+\ell})}}
{\sum_{\tilde{x}^\ell\in\calX^\ell}
2^{-LZ(\tilde{x}^\ell)}}\nonumber\\
&\ge&\prod_{i=0}^{n/\ell-1}2^{-LZ(\hat{x}_{\ell
i+1}^{\ell i+\ell})}\nonumber\\
&=&\exp_2\left\{-\sum_{i=0}^{n/\ell-1}LZ(\hat{x}_{\ell i+1}^{\ell
i+\ell})\right\}\nonumber\\
&\ge&\exp_2\left\{-\sum_{i=0}^{n/\ell-1}c(\hat{x}_{\ell i+1}^{\ell i+\ell})\log
c(\hat{x}_{\ell i+1}^{\ell i+\ell})-n\cdot\epsilon(\ell)\right\},
\end{eqnarray}
where the last step follows from eq.\ (\ref{lz-clogc}) applied to
$\ell$--vectors.
It remains to show that
$\sum_{i=0}^{n/\ell-1}c(\hat{x}_{\ell i+1}^{\ell i+\ell})\log
c(\hat{x}_{\ell i+1}^{\ell i+\ell})$ is essentially no larger than
$\rho_K(\hat{x}^n)$ for some $K$ that can be chosen arbitrarily large,
provided that $n \gg \ell$ and $\ell$ is large enough. Consider the following chain of
inequalities for a given positive integer $K$:
\begin{eqnarray}
& &\sum_{i=0}^{n/\ell-1}\{c(\hat{x}_{\ell i+1}^{\ell i+\ell})\log
c(\hat{x}_{\ell i+1}^{\ell i+\ell})-[c(\hat{x}_{\ell i+1}^{\ell
i+\ell})+K^2]\log(4K^2)\}\nonumber\\
&\le&\ell\sum_{i=0}^{n/\ell-1}\rho_K(\hat{x}_{\ell i+1}^{\ell i+\ell})\nonumber\\
&=&\ell\cdot\sum_{i=0}^{n/\ell-1}\min_{E\in\calE(K)}
\rho_E(\hat{x}_{\ell i+1}^{\ell i+\ell})\nonumber\\
&=&\sum_{i=0}^{n/\ell-1}\min_{E\in\calE(K)}\sum_{j=1}^\ell l[p(\sigma_{\ell
i+j},\hat{x}_{\ell
i+j})]\nonumber\\
&\le&\min_{E\in\calE(K)}\sum_{i=0}^{n/\ell-1}\sum_{j=1}^\ell 
l[p(\sigma_{\ell i+j},\hat{x}_{\ell
i+j})]\nonumber\\
&=&\min_{E\in\calE(K)}\sum_{i=1}^nl[p(\sigma_i,\hat{x}_i)]\nonumber\\
&=&n\cdot\rho_K(\hat{x}^n),
\end{eqnarray}
where the first inequality follows from \cite[Theorem 1]{ZL78} applied to
$\ell$--vectors. Thus,
\begin{eqnarray}
\sum_{i=0}^{n/\ell-1}c(\hat{x}_{\ell i+1}^{\ell i+\ell})\log c(\hat{x}_{\ell
i+1}^{\ell i+\ell})&\le&
n\cdot\rho_K(\hat{x}^n)+\sum_{i=0}^{n/\ell-1}[c(\hat{x}_{\ell
i+1}^{\ell i+\ell})+K^2]\log(4K^2)\nonumber\\
&\le&n\cdot\rho_K(\hat{x}^n)+\frac{n}{\ell}\cdot\left[\frac{\ell\log(4K^2)}
{(1-\epsilon_\ell)\log\ell}+K^2\log(4K^2)\right]\nonumber\\
&=&n\cdot\left[\rho_K(\hat{x}^n)+\frac{\log(4K^2)}{(1-\epsilon_\ell)\log\ell}+
\frac{K^2\log(4K^2)}{\ell}\right],
\end{eqnarray}
and so,
\begin{eqnarray}
\sum_{i=0}^{n/\ell-1}LZ(\hat{x}_{\ell i+1}^{\ell i+\ell})&\le&
\sum_{i=0}^{n/\ell-1}c(\hat{x}_{\ell
i+1}^{\ell i+\ell})\log c(\hat{x}_{\ell
i+1}^{\ell i+\ell})+n\epsilon(\ell)\nonumber\\
&\le&n\cdot\left[\rho_K(\hat{x}^n)+\frac{\log(4K^2)}{(1-\epsilon_\ell)\log\ell}+
\frac{K^2\log(4K^2)}{\ell}+\epsilon(\ell)\right].
\end{eqnarray}
This completes the proof of Theorem \ref{directtheorem}.

\section{Side Information}
\label{si}

We now consider the extended setting where a deterministic side
information vector, $y^n$, is available to the randomized guessing machine. 
Since most of the ideas and techniques extend quite straightforwardly, we only outline the
differences compared to the case without side information.

We now define the model as follows.
An FSGM is defined by a set
$Q=(\calU,\calX,\calY,\calZ,\ell,f,g,\Delta)$,
where $\calU$, 
$\calX$, and $\calZ$ are as before,
$\calY$ is the finite alphabet of size $\beta$ associated with the side
information, $\ell$ is a positive integer,
$f:\calZ\times\calY^\ell\times\calU^*\to\calX^\ell$ is the
output function, $g:\calZ\times\calY^\ell\times\calU^*\to\calZ$ is the
next--state function, and
$\Delta:\calZ\times\calY^\ell\to \{0,1,2,\ldots\}$.
When $\bu=u_1,u_2,\ldots$
and the side information sequence, $\by=y_1,y_2,\ldots$, $y_t\in\calY$, $t=1,2,\ldots$, are fed
into $Q$, it produces 
$\hat{\bx}^n$, according to
\begin{eqnarray}
t_i&=&t_{i-1}+\Delta(z_i,y_{(i-1)\ell+1}^{i\ell}),~~~~~~t_0\dfn 0 \label{ti1}\\
v_i&=&(u_{t_{i-1}+1},u_{t_{i-1}+2},\ldots,u_{t_i}), \label{ki1}\\
\hat{x}_{(i-1)\ell+1}^{i\ell}&=&f(z_i,y_{(i-1)\ell+1}^{i\ell},v_i), \label{yi1}\\
z_{i+1}&=&g(z_i,y_{(i-1)\ell+1}^{i\ell},v_i). \label{nextstate1}
\end{eqnarray}
Note that here, we have somewhat generalized the model in the sense that the system is
now fed by $\ell$--tuples of $\by$ and it produces $\ell$--tuples of $\hat{\bx}$.
The reason is that in the context of systems with a side information input,
input--output mechanisms that work on a symbol--by--symbol basis
(i.e., $\ell=1$) are too limited. It is reasonable to 
allow dependencies between side information symbols and their
corresponding output symbols with some delay and anticipation, and indeed, 
such a delay and anticipation will be needed in the
achievability scheme.
We could have allowed a general $\ell$
also in the earlier case, where no side information was available.

Let $G_Q(x^n|y^n)$ denote the random number of guessing rounds
needed for $Q$ until success.
Next, for a given $\zeta > 0$, define
\begin{equation}
\gamma_{s,\ell}(x^n|y^n)=\min_{Q\in\calQ(s,\ell)}\frac{\ln\bE\{[G_Q(x^n|y^n)]^\zeta\}}{n},
\end{equation}
where $\calQ(s,\ell)$ is the set of all FSGMs with block length less than or
equal to $\ell$
and no more than $s$ states.
For two given infinite sequences, $\bx=(x_1,x_2,\ldots)$ and
$\by=(y_1,y_2,\ldots)$, we define
\begin{equation}
\gamma_{s,\ell}(\bx|\by)=\limsup_{n\to\infty}\gamma_{s,\ell}(x^n|y^n),
\end{equation}
and finally,
\begin{equation}
\gamma(\bx|\by)=\lim_{s\to\infty}\lim_{\ell\to\infty}\gamma_{s,\ell}(\bx|\by).
\end{equation}

To present the results, we need a few more definitions. 
Consider the joint parsing of the sequence of pairs,
$\{(x_1,y_1),(x_2,y_2),\ldots,(x_n,y_n)\}$, let $c(x^n,y^n)$ denote
the number of phrases, $c(y^n)$ -- the number of distinct $y$-phrases,
$\by(j)$ -- the $j$-th distinct $y$-phrase, $1\le j\le c(y^n)$, and
finally, let
$c_j(x^n|y^n)$ denote the number of times
$\by(j)$ appears as a phrase, or, equivalently,
the number of distinct $x$-phrases that appear jointly with $\by(j)$,
so that $\sum_{j=1}^{c(y^n)}c_j(x^n|y^n)=c(x^n,y^n)$. Then, we define
the conditional LZ complexity \cite{Ziv85} as
\begin{equation}
\label{def. LZ(x|y)}
u(x^n,y^n)=\sum_{j=1}^{c(y^n)}c_j(x^n|y^n)\log c_j(x^n|y^n).
\end{equation}
Let the conditional $K$--state compressibility of $x^n$ given $y^n$, denoted
$\rho_K(x^n|y^n)$, be defined
as in \cite[pp.\ 2245]{merhav00}: A $K$-state
encoder $E$ with side information is defined by a set of six objects
$(\Sigma,\calB,\calX,\calY,p,q)$, where
$\Sigma$ is a finite set
of $K$ states, $\calB$ is
a finite set of binary words
(possibly of different lengths, including the null word for idling),
$\calX$ is the finite
alphabet of the source to be compressed,
$\calY$ is a finite alphabet
of side information,
$p:\Sigma\times\calX\times\calY\to\calB$ is the encoder output function, and
$q:\Sigma\times\calX\times\calY\to\Sigma$ is the next--state function.
When an input sequence $x_1,x_2,\ldots$ and a side information
sequence $y_1,y_2,\ldots$ are fed
together, sequentially into $E$, the
encoder outputs a sequence of binary words $b_1,b_2,\ldots$, $b_i\in\calB$,
according to
\begin{equation}
\label{fsmencoder}
b_i=p(\sigma_i,x_i,y_i),~~~\sigma_{i+1}=q(\sigma_i,x_i,y_i),~~~i=1,2,...
\end{equation}
where $\sigma_i$ is the state of $E$ at time instant $i$.
The decoder, on the other hand, receives the pair sequence
$(b_1,y_1),(b_2,y_2),\ldots$
and reconstructs the source sequence $x_1,x_2,\ldots$.
A finite--state encoder $E$
with side information is said to be {\it information
lossless} (IL) if for all $\sigma_1\in\Sigma$ and all
$(x^n,y^n)\in\calX^n\times\calY^n$, $n\ge 1$,
the quadruple $(\sigma_1,\sigma_{n+1},b^n,y^n)$ uniquely determines $x^n$,
The length function
associated with $E$ is defined as $L_E(x^n|y^n)=\sum_{i=1}^nl(b_i)$,
where $l(b_i)$ is the length of the binary string $b_i\in\calB$.
We now define 
\begin{equation}
\rho_K(x^n|y^n)=\min_{E\in\calE(K)}\frac{L_E(x^n|y^n)}{n}
\end{equation}
where $\calE(K)$ is the class of IL encoders with no more than $K$ states.
As shown in \cite[eq.\ (13)]{merhav00},
\begin{eqnarray}
\rho_K(x^n|y^n)&\ge&\sum_{j=1}^{c(y^n)}[c_j(x^n|y^n)+K^2]\log
\frac{c_j(x^n|y^n)+K^2}{4K^2}\nonumber\\
&\ge &u(x^n,y^n)-[c(x^n,y^n)+K^2]\log(4K^2)\nonumber\\
&\ge&u(x^n,y^n)-\frac{n\log(4K^2)}{(1-\epsilon_n)\log n}-K^2\log(4K^2).
\end{eqnarray}

\subsection{Converse Bounds} 

The converse bounds are as follows.
\begin{eqnarray}
\gamma_{s,\ell}(x^n|y^n)&\ge&\zeta\cdot
\left[\rho_{K(\ell))}(x^n|y^n)-
\frac{\log(2s^3e)}{\ell}\right]-\frac{\log(e^22^\zeta)}{n}\\
\gamma_{s,\ell}(x^n|y^n)&\ge&\zeta\bigg[u(x^n,y^n)-n\delta_n(s,\ell)\bigg],
\end{eqnarray}
where now $K(\ell)$ is redefined as $([\alpha\beta]^{k+1}-1)/(\alpha\beta-1)$
and $\delta_n(s,\ell)$ tends to zero uniformly as $n\to\infty$ for any fixed
$(s,\ell)$. Note that now there is no maximization over all values of $\ell$
that are divisors of $n$ because now $\ell$ is a parameter of the model and
not an auxiliary parameter as before (indeed,
$\gamma_{s,\ell}(x^n|y^n)\gamma_{s,\ell}(x^n|y^n)$ at the l.h.s.\ depends on
$\ell$ too).

The proof follows the same lines as before with a just few twists.
Any FSGM $Q$ with $s$ states incudes a channel from $y^n$ to $x^n$ with the
following structure,
\begin{equation}
P(\hat{x}^n|y^n)=\sum_{z_2,z_3,\ldots,z_{n+1}}\prod_{i=1}^{n/\ell}
P(\hat{x}_{(i-1)\ell+1}^{i\ell},z_{i+1}|z_i,y_{(i-1)\ell+1}^{i\ell}).
\end{equation}
As in the earlier derivation, we have
\begin{equation}
\label{guess-prob-si}
\bE\{[G_Q(x^n|y^n)]^\zeta\}\ge
\frac{2^{-\zeta}}{e^2}\cdot \exp\{-\zeta\ln P(x^n|y^n)\}.
\end{equation}
Consider now the partitioning $\hat{x}^n$
and $y^n$ into
$m=n/\ell$ non--overlapping segments of length $\ell$. Then, once again,
\begin{equation}
\label{prob-ent-si}
-\log P(\hat{x}^n|y^n)\ge m[\hat{H}_\ell(x^n|y^n)-\log(s^3e)],
\end{equation}
where $\hat{H}_\ell(x^n|y^n)$ is the conditional empirical entropy of
$\ell$--blocks. Similarly as in the earlier derivation, 
we can further lower bound the r.h.s.\ in terms of 
the conditional compressibility.
\begin{equation}
n\rho_{K(\ell)}(\hat{x}^n|y^n)\le m\hat{H}_\ell(\hat{x}^n|y^n)+m,
\end{equation}
and combining this with eq.\ (\ref{prob-ent-si}), we obtain
\begin{equation}
-\log P(\hat{x}^n|y^n)\ge n\rho_{K(\ell)}(\hat{x}^n|y^n)-m\log(2s^3e),
\end{equation}
which, together with eq.\ (\ref{guess-prob-si}),
proves the first converse bound, and
the second lower bound follows from \cite{merhav00}
\begin{eqnarray}
n\rho_{K(\ell)}(\hat{x}^n|y^n)&\ge&\sum_{j=1}^{c(y^n)}[c_j(\hat{x}^n|y^n)+K^2(\ell)]\log
\frac{c_j(\hat{x}^n|y^n)+K^2(\ell)}{4K^2(\ell)}\nonumber\\
&\ge& \sum_{j=1}^{c(y^n)}c_j(\hat{x}^n|y^n)\log c_j(\hat{x}^n|y^n)-
[c(\hat{x}^n,y^n)+K^2(\ell)]\log[4K^2(\ell)]\nonumber\\
&\ge& u(\hat{x}^n,y^n)-
\frac{n\log[4K^2(\ell)]}{(1-\epsilon_n)\log n}-K^2(\ell)\log[4K^2(\ell)].
\end{eqnarray}
It follows that the second converse bound holds with
\begin{equation}
\delta_n(s,\ell)=\frac{\log[4K^2(\ell)]}{(1-\epsilon_n)\log
n}+\frac{K^2(\ell)\log[4K^2(\ell)]}{n}+\frac{\log(2s^3e)}{\ell}+\frac{\log(e^22^\zeta)}{n}.
\end{equation}

\subsection{Achievability}

Following the same steps as in Section \ref{direct} and in \cite{MC19}, 
consider randomly drawing guesses according to the
distribution
\begin{equation}
P(\hat{x}^n|y^n)=\frac{2^{-LZ(\hat{x}^n|y^n)}}{\sum_{\tilde{x}^n}
2^{-LZ(\hat{x}^n|y^n)}},
\end{equation}
where $LZ(\hat{x}^n|y^n)$ is the length of compressed version of
$\hat{x}^n$ given $y^n$ using the conditional version of the
LZ78 algorithm \cite[p.\ 460]{Ziv85} (see also \cite{UK03}).
It is easy to see that this randomized guessing distribution 
asymptotically achieves the second lower bound, since
\begin{equation}
LZ(\hat{x}^n|y^n)\le u(x^n,y^n)+n\epsilon_1(n),
\end{equation}
where $\epsilon_1(n)$ is of the order of $\log(\log n)/(\log n)$,
as shown in \cite[p.\ 460]{Ziv85}.
Once again, to devise a matching direct in the framework of finite--state
machines, we can restart every $\ell$--block and apply the random guessing
distribution
\begin{eqnarray}
P(\hat{x}^n|y^n)&=&\prod_{i=0}^{n/\ell-1}\left[\frac{2^{-LZ(\hat{x}_{\ell
i+1}^{\ell i+\ell}|y_{\ell i+1}^{\ell i+\ell})}}
{\sum_{\tilde{x}^\ell\in\calX^\ell}
2^{-LZ(\tilde{x}^\ell|y_{\ell i+1}^{\ell i+\ell})}}\right]\nonumber\\
&\ge&\exp_2\left\{-\sum_{i=0}^{n/\ell-1}u(\hat{x}_{\ell i+1}^{\ell
i+\ell}, y_{\ell i+1}^{\ell i+\ell})-n\epsilon_1(\ell)
\right\}.
\end{eqnarray}
We now need to show that
$\sum_{i=0}^{n/\ell-1}u(\hat{x}_{\ell i+1}^{\ell i+\ell},y_{\ell i+1}^{\ell
i+\ell})$ is 
essentially no larger than
$\rho_K(\hat{x}^n|y^n)$ for some $K$ that can be chosen arbitrarily large,
provided that $n$ is large enough. Once again, consider the following chain of
inequalities for a given positive integer $K$:
\begin{eqnarray}
& &\sum_{i=0}^{n/\ell-1}\sum_{j=1}^{c(y_{\ell i+1}^{\ell i+\ell})}
[c_j(\hat{x}_{\ell i+1}^{\ell i+\ell}|y_{\ell i+1}^{\ell i+\ell})+K^2]\log
\left[\frac{c_j(\hat{x}_{\ell i+1}^{\ell i+\ell}|y_{\ell i+1}^{\ell
i+\ell})+K^2}{4K^2}\right]\nonumber\\
&\le&\ell\sum_{i=0}^{n/\ell-1}\rho_K(\hat{x}_{\ell i+1}^{\ell
i+\ell}|y_{\ell i+1}^{\ell i+\ell})\nonumber\\
&=&\ell\cdot\sum_{i=0}^{n/\ell-1}\min_{E\in\calE(K)}
\rho_E(\hat{x}_{\ell i+1}^{\ell i+\ell}|y_{\ell i+1}^{\ell i+\ell}))\nonumber\\
&=&\sum_{i=0}^{n/\ell-1}\min_{E\in\calE(s)}\sum_{j=1}^\ell l[p(\sigma_{\ell
i+j},\hat{x}_{\ell
i+j},y_{\ell i +j})]\nonumber\\
&\le&\min_{E\in\calE(K)}\sum_{i=0}^{n/\ell-1}\sum_{j=1}^\ell 
l[p(\sigma_{\ell i+j},\hat{x}_{\ell
i+j},y_{\ell i+j})]\nonumber\\
&=&\min_{E\in\calE(K)}\sum_{i=1}^nl[p(\sigma_i,\hat{x}_i,y_i)]\nonumber\\
&=&n\cdot\rho_K(\hat{x}^n|y^n).
\end{eqnarray}
Thus,
\begin{eqnarray}
\sum_{i=0}^{n/\ell-1}
\sum_{j=1}^{c(y_{\ell i+1}^{\ell i+\ell})}
c_j(\hat{x}_{\ell i+1}^{\ell i+\ell}|y_{\ell i+1}^{\ell i+\ell})\log c_j(\hat{x}_{\ell
i+1}^{\ell i+\ell}|y_{\ell i+1}^{\ell i+\ell})&\le&
n\cdot\rho_K(\hat{x}^n|y^n)+\sum_{i=0}^{n/\ell-1}
\sum_{j=1}^{c(y_{\ell i+1}^{\ell i+\ell})}
c_j(\hat{x}_{\ell
i+1}^{\ell i+\ell}|y_{\ell i+1}^{\ell i+\ell})\log(4K^2)\nonumber\\
&\le&n\cdot\rho_K(\hat{x}^n|y^n)+\frac{n}{\ell}\cdot\frac{\ell\log(\alpha\beta)\cdot\log(4K^2)}
{(1-\epsilon_\ell)\log\ell}\nonumber\\
&=&n\cdot\left[\rho_K(\hat{x}^n|y^n)+\frac{\log(\alpha\beta)\log(4K^2)}
{(1-\epsilon_\ell)\log\ell}\right]
\end{eqnarray}
and the remaining steps are similarly as before.

\end{document}